\documentclass[prc,twocolumn,showpacs,superscriptaddress,floatfix]{revtex4}
\usepackage{graphicx}

\def\be{\begin{equation}}
\def\ee{\end{equation}}
\def\ba{\begin{eqnarray}}
\def\ea{\end{eqnarray}}
\def\bas{\begin{eqnarray*}}
\def\eas{\end{eqnarray*}}

\begin{document}

\title{Energy functional for the three-level Lipkin model}

\author{Michael G. Bertolli} 

\affiliation{Department of Physics and Astronomy, University of
Tennessee, Knoxville, TN 37996, USA}

\author{Thomas Papenbrock}

\affiliation{Department of Physics and Astronomy, University of
Tennessee, Knoxville, TN 37996, USA}

\affiliation{Physics Division, Oak Ridge National Laboratory, Oak
Ridge, TN 37831, USA}

\date{\today}

\begin{abstract}
We compute the energy functional of a three-level Lipkin model via a
Legrendre transform and compare exact numerical results with
analytical solutions obtained from the random phase approximation
(RPA). Except for the region of the phase transition, the RPA
solutions perform very well.  We also study the case of three
non-degenerate levels and again find that the RPA solution agrees well
with the exact numerical result. For this case, the analytical results
give us insight into the form of the energy functional in the presence
of symmetry-breaking one-body potentials.
\end{abstract}


\maketitle

\section{Introduction}

Nuclear density functional theory (DFT) is applicable for a
description of ground-state properties throughout the nuclear
chart~\cite{Gor02,Lun03,Ben03,Sto03}. This theory is based on the
theorem by Hohenberg and Kohn~\cite{Hoh64}, and its practical
applications are based on the self-consistent mean-field theory with
density-dependent energy
functionals~\cite{Sk56,Koh65,Vau72,Vau73}. 
Within nuclear DFT, the solution of the quantum nuclear many-body
problem is relatively simple. The problem is, of course, with the
construction of the energy functional.

Most approaches to the nuclear energy-density functional are empirical
in nature, and the inclusion of pairing
properties\cite{Dob84,Oli88,Rob01,Dug01,Furn06,Dug07} is particularly
important and challenging. There are only a few non-empirical
approaches to the nuclear energy-density functional. The
density-matrix expansion pioneered by Negele and
Vautherin~\cite{Neg72} expands the non-diagonal one-body density
matrix along its diagonal, and the expectation value of a Hamiltonian
thus leads to an energy-density functional.  In another formal
approach~\cite{Lieb83}, the energy-density functional is constructed
as the Legendre transform of the ground-state energy as a functional
of external one-body potentials. This approach is closely related to
the Hohenberg-Kohn theorem~\cite{Hoh64}. However, its starting point,
i.e. the ground-state energy of a quantum many-body system as a
functional of any external potential, is only available for a few
solvable or weakly interacting systems. Furnstahl and
co-workers~\cite{Pug03,Bhat05} followed this path and derived
energy-density functionals for dilute Fermi gases with short-ranged
interactions. For Fermi gases in the unitary regime, simple scaling
arguments suggest the form of the energy density
functional~\cite{Car03,Pap05,Bhat06,Bul07}.

In nuclear physics, energy-density functional theory is a practical
tool that is popular due to its computational simplicity and success.
The universality of the functional, i.e. the possibility to study
nucleons in external potentials, is seldom used. This distinguishes
nuclear DFT from DFT in Coulomb systems and makes alternative
formulations worth studying. For computational simplicity, we would
like to maintain the framework of an energy functional. However,
there is no need to focus on functionals of the density. The
description and interpretation of nuclear systems is often based on
shell-model orbitals rather than densities. For instance, in nuclear
structure, one is often more interested in the occupation of a given
shell-model orbital or the isospin-dependence of the effective
single-particle energies than in the shape of the density
distribution. This language is natural for shell-model Hamiltonians
that are based on single-particle orbitals. It is the purpose of this
paper to study and construct the energy functional of such a system.

Recently, the energy functional (in terms of occupation numbers) was
constructed for the pairing Hamiltonian~\cite{Pap07}. In this paper we
consider the three-level Lipkin model~\cite{LMG,Li,Yukawa,Meredith}. This
is another solvable model that is relevant for nuclei. We will be able
to gain insight into the formulations of energy functionals for
two-body interactions that exhibit a continuous symmetry, and will
study the effect of symmetry-breaking one-body potentials.

This paper is organized as follows. In Sect.~\ref{degen} we construct
the energy functional of the three-level Lipkin model with degenerate
excited levels. We study the implications of symmetry-breaking
one-body potentials in Sect.~\ref{symbreak}, and conclude with our
Summary.

\section{Degenerate energy levels}
\label{degen}
We consider the three-level Lipkin model~\cite{LMG,Li,Yukawa,Meredith}
and follow the work by Hagino and Bertsch~\cite{HaginoBertsch}. This
model consists of $N$ fermions that are distributed over three levels
(labeled as 0, 1, and 2) consisting of $N$ degenerate, single-particle
states each.  The single-particle energies are $\varepsilon_0=0$, and 
the two excited levels are degenerate,
$\varepsilon_1 =\varepsilon_2\equiv \varepsilon$.  The Hamiltonian is
\ba
H&=&\varepsilon(\hat{n}_1+\hat{n}_2) \nonumber\\
&&-\frac{V}{2}(K_1K_1+K_2K_2+K^\dagger_1K^\dagger_1+K^\dagger_2K^\dagger_2)
\ .
  \label{hamiltonian}
\ea
Here,
\be
  \hat{n}_\alpha=\sum_{i=1}^N c^\dagger_{\alpha i}c_{\alpha i},\quad \alpha=0,1,2 
\ee
is the number operator of level $\alpha$, while
\be
 K_\alpha=\sum_{i=1}^{N} c^\dagger_{\alpha i}c_{0 i},\quad \alpha=1,2 
\ee  
transfer fermions from the level 0 to the level $\alpha>0$. We assume
$\varepsilon\ge 0$ for the spacing to the degenerate levels 1 and 2.
The operators $c_{\alpha i}^\dagger$ and $c_{\alpha i}$ create and
annihilate a fermion in state $i$ of level $\alpha$.  The Hamiltonian
is invariant under the simultaneous orthogonal transformation of the
$N$ orbitals belonging to each level, and this symmetry facilitates
the numerical solution of this problem~\cite{Meredith}. Note that the
Hamiltonian~(\ref{hamiltonian}) is invariant under orthogonal
transformations of states belonging to the degenerate levels 1 and 2,
i.e.
\ba
\label{symmetry}
\left(\begin{array}{c}
c_{1 i}^\dagger\\
c_{2 i}^\dagger
\end{array}\right) \to
\left(\begin{array}{rr}
\cos{\phi} & \sin{\phi}\\
-\sin{\phi}& \cos{\phi}
\end{array}\right)
\left(\begin{array}{c}
c_{1 i}^\dagger\\
c_{2 i}^\dagger
\end{array}\right) \ .
\ea
This symmetry will be broken by the Hartree-Fock (HF) solution (see
below), and we will study the explicit breaking of this symmetry by
additional one-body terms of the Hamiltonian in Sect.~\ref{symbreak}.

\begin{figure}
  \includegraphics[width=0.35\textwidth,clip=]{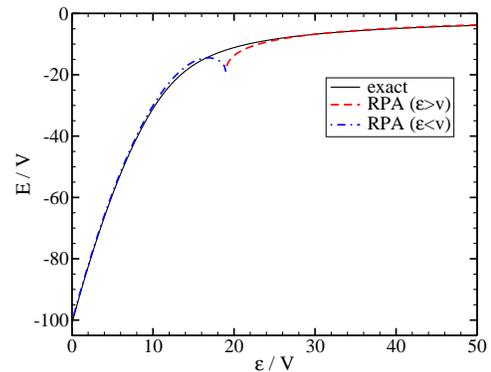}
  \caption{(color online) Comparison of the exact ground-state energy
  to the RPA approximation for a system of $N=20$ fermions. The RPA
  result shows a cusp at $\varepsilon=v\equiv (N-1)V$, corresponding
  to the boundary between the spherical $(\varepsilon>v)$ and the
  deformed phase $(\varepsilon <v)$
 \label{RPACusp}}
\end{figure}

The energy functional is the Legendre transform
\begin{equation}
  F(n)=E(\varepsilon)-  \varepsilon n(\varepsilon) \ . 
  \label{FuncExact1}
\end{equation}
Here 
\be
\label{nocc}
n(\varepsilon)\equiv {\partial E(\varepsilon)\over\partial \varepsilon}
\ee 
is the ground-state occupation of levels 1 and 2, while
$E(\varepsilon)$ is the ground-state energy as a function of the
spacing $\varepsilon$. The exact ground-state energy can be obtained
numerically by diagonalizing a matrix of modest
dimensions~\cite{Meredith}, and one can thus construct easily the
exact energy functional. However, we are interested in gaining
analytical insight into the problem. For this reason, we employ the
analytical expression by Hagino and Bertsch~\cite{HaginoBertsch} that
expresses the ground-state energy in terms of the HF
energy and corrections due to the RPA.
The HF result depends on the size of the parameter $\varepsilon/v$
where
\begin{equation}
  v\equiv V(N-1) 
\end{equation}
is the effective strength of the two-body coupling $V$ in the $N$-body
system.  The spherical phase is found for $\varepsilon/v>1$, while one
deals with a deformed phase for $\varepsilon/v<1$. In the spherical
(deformed) phase, the HF solution preserves (breaks) the
symmetry~(\ref{symmetry}) of the Hamiltonian.  The results
are~\cite{HaginoBertsch}
\begin{eqnarray}
\label{erg1}
  E&=& 
  \sqrt{\varepsilon^2-v^2}-\varepsilon\quad\mbox{for}\quad\varepsilon>v \ ,\\
\label{erg2}
  E&=& 
  -\frac{N+4}{4}v + \sqrt{{v^2-\varepsilon^2\over 2}} \nonumber\\
  &&+{2N-1\over 4}\varepsilon - {N-1\over 4}{\varepsilon^2\over v}\quad\mbox{for}\quad\varepsilon<v \ .
\end{eqnarray}
Figure~\ref{RPACusp} compares the approximate ground-state
energies~(\ref{erg1}) and (\ref{erg2}) with the exact result as a
function of $\varepsilon/V$ for a system of $N=20$ fermions. The
approximate solution exhibits a cusp at the boundary $\varepsilon/v=1$
between the spherical and deformed phases, but it agrees well with the
exact result away from the phase boundary. Note that the cusp renders
the RPA solution non-convex, and its Legendre transform is therefore not
possible close to the phase boundary. Within the RPA formalism, the
occupation number~(\ref{nocc}) attains non-physical negative values
close to the cusp. We will avoid this problem in the construction presented 
below.

Let us start with the analytical construction of the energy 
functional~(\ref{FuncExact1}) in the spherical phase. Here, the RPA solution 
is convex. We have
\begin{equation}
  n\equiv\frac{\partial E}{\partial \varepsilon} = \frac{\varepsilon}{\sqrt{\varepsilon^2-v^2}}-1\ .
  \label{occnumspherical}
\end{equation}
This equation can easily be solved for $\varepsilon(n)$, and the
functional thus reads
\begin{equation}
\label{fsym1}
  F(n)=-|v|\sqrt{n(n+2)} \ .
\end{equation}

Thus, for small occupation numbers (i.e. the weak coupling limit $v
\ll \varepsilon$), the functional is nonanalytical and exhibits a
square root singularity. This is a rather generic feature of energy
functionals and is also seen in the energy functional for the pairing
Hamiltonian.
 
We next turn to the deformed case $(\varepsilon< v)$.  The occupation number is
\be
  n(\varepsilon)= \frac{2N-1}{4}-{N-1\over 2} {\varepsilon\over v}-   {\varepsilon\over \sqrt{2(v^2-\varepsilon^2)}} \ .
  \label{occnumdeformed}
\ee
Equation~(\ref{occnumdeformed}) is difficult to invert analytically
because of the term involving the square root. Note that this term
also renders the occupation number negative as $\varepsilon$
approaches $v$. We avoid this problem by expanding
Eq.~(\ref{occnumdeformed}) in powers of $\varepsilon/v$, approximate
$\varepsilon/v\ll 1$, and keep up to orders $O(\varepsilon/v)$ in the
resulting expression.  The result is
\begin{equation}
\varepsilon(n)=\frac{N-2n-1/2}{ N-1+\sqrt{2}} v \ .
\end{equation}
Insertion of this result and Eq.~(\ref{erg2}) into
Eq.~(\ref{FuncExact1}) yields the density functional in the deformed
phase.

Note that the energy functional is analytical in the strong coupling
limit $v\gg\varepsilon$. Again, this is a generic feature of energy
functionals and was also exhibited in the pairing problem.

\begin{figure}
  \includegraphics[width=0.35\textwidth,clip=]{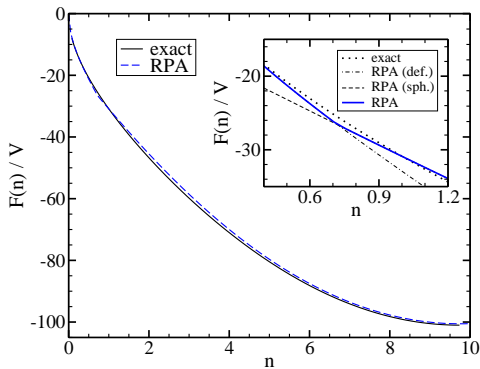}
  \caption{(color online) Comparison of the exact energy functional
  (solid line) and RPA-obtained functional (dashed line) for $N=20$
  fermions.  Inset: The phase transition is clearly seen at the
  intersection of the deformed and spherical RPA solutions.
  \label{FuncFig}}
\end{figure}

Figure~\ref{FuncFig} compares the exact energy functional with the RPA
solutions obtained in the spherical and deformed phase for $N=20$.
Note that the spherical (deformed) phase corresponds to sufficiently
small (large) occupation numbers. The inset shows that the RPA
solution has a kink at the critical occupation number $n \approx 0.7$
at the phase transition. Note also that the solution of
the deformed phase is a good approximation over the entire range $n>0.7$,
although its derivation employed the approximation $\varepsilon\ll v$
corresponding to $n\approx N/2$, see Eq.~(\ref{occnumdeformed}). 

How severe is the presence of the kink at $n\approx 0.7$ in the
functional?  To address this question, we employ the RPA functional in
practical calculations. We add the one-body term $\varepsilon n$ to the
functional and numerically determine the occupation number $n$ that minimizes
the ground-state energy 
\begin{equation}
  E(n)=F(n)+\varepsilon n \ .
\end{equation}

\begin{figure}
  \includegraphics[width=0.35\textwidth,angle=0]{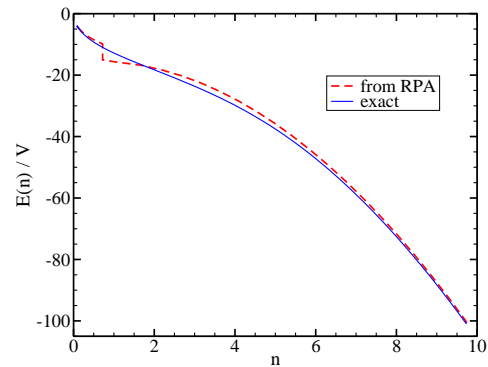}
  \caption{(color online) Ground-state energy as a function of
  occupation number obtained from the exact functional (solid line)
  and the RPA functional (dashed line) for $N=20$ fermions. The RPA
  solution fails at the phase transition. \label{FuncTest}}
\end{figure}

The energy taken at the minimum is plotted as a function of the
minimizing occupation number $n$ in Fig.~\ref{FuncTest}.  The result
is also compared to the exact numerical result.  We see that the
energy functional, obtained via RPA, provides a good prediction of the
ground state energy function of the three-level Lipkin model.  At the
phase transition, however, the functional predicts multiple values for
the ground state energy, due to the discontinuity in the derivative at
the phase transition.  Thus, for external potentials
$\varepsilon\approx v$ the energy functional does not provide good
predictive power, and the relatively small kink visible in the inset
of Fig.~\ref{FuncFig} is, of course, the reason for this shortcoming.

\section{Non-degenerate energy levels and symmetry-breaking}
\label{symbreak}

The two-body interaction and the one-body terms of the
Hamiltonian~(\ref{hamiltonian}) are invariant under the symmetry
transformation~(\ref{symmetry}). In this section we study the breaking
of this symmetry by a one-body potential, i.e. we lift the degeneracy
of levels 1 and 2 in the Hamiltonian~(\ref{hamiltonian}) and modify
its one-body term as follows
\begin{equation}
    \varepsilon (\hat{n}_1+\hat{n}_2)\to  \varepsilon_1 \hat{n}_1+\varepsilon_2 \hat{n}_2 \ .
\end{equation}
The two-body interaction remains the same. Technically it is
convenient to substitute
$\varepsilon\to(\varepsilon_1+\varepsilon_2)/2$ in
Eq.~(\ref{hamiltonian}) and to add the term
\begin{equation}
\frac{\varepsilon_1-\varepsilon_2}{2}(\hat{n}_1-\hat{n}_2)
\label{NondegenerateAddition}
\end{equation}
to the resulting expression. This approach is particularly convenient
for the RPA calculation.

The question of symmetry-breaking is an important one. The energy
functional corresponds to the symmetry-preserving two-body interaction
that is probed by a symmetry-breaking one-body potential. Thus, the
functional depends on two variables (the occupation numbers of levels 1
and 2), and we are interested in the form of this functional and in
its relation to the functional of the symmetry-preserving case. We
thereby hope to get insights into energy functionals for atomic
nuclei.  The nuclear shell model, for instance, employs a spherically
symmetric two-body interaction, but deformed nuclei are often
described within a deformed mean-field basis that breaks the
rotational invariance through one-body terms. 

The numerical calculations are simple and require only a minor
modification of the matrix elements corresponding to the one-body
terms. For the analytical construction of the functional, we have to
perfom the HF and RPA calculations. We adapt the HF calculation of
Ref.~\cite{HaginoBertsch} to the case of non-degenerate levels and 
obtain the HF-energy surface
\begin{eqnarray}
  E(\alpha,\beta)&=&N 
\sin^2 \alpha(\varepsilon_1 \cos^2\beta+\varepsilon_2\sin^2\beta) \nonumber \\
  &&-vN\sin^2\alpha \cos^2\alpha
  \label{EnTransform}
\end{eqnarray}
as a function of the two angles $\alpha$ and $\beta$ of the HF
transformation.  In the case $\varepsilon_1=\varepsilon_2$ of
degenerate levels, the energy surface~(\ref{EnTransform}) becomes
independent of $\beta$.

In what follows, we assume $\varepsilon_1\le\varepsilon_2$.  The
minimum of the energy surface~(\ref{EnTransform}) occurs at $\alpha=0$
(where $\beta$ is arbitrary and can be chosen as zero) for
$\varepsilon_1>v$, and at $\cos 2\alpha=\varepsilon_1/v$ and $\beta=0$
for $\varepsilon_1<v$.  Again, we refer to these cases as the
weak-coupling regime and the strong-coupling regime, respectively.  
The Hartree-Fock energy thus becomes
\begin{equation}
E_{{\rm HF}}=\left\{
  \begin{array}{l l}
     0\ , &\varepsilon_1>v\\
     -{N\over 4}v +{N\over 2}\varepsilon_1-{N\over 4}\varepsilon_1^2/v\ ,& 
\varepsilon_1<v
  \end{array} \right. \ .
\end{equation}

Thus, within the HF approximation, all particles stay in level 0 in the
weak-coupling regime, while the fermions occupy level 0 and level 1 in the
strong-coupling regime. The RPA equation takes the well-known form
\begin{eqnarray}
  \left(
    \begin{array}{c c}
      A&B \\
      -B&-A
    \end{array}
  \right)
  \left(
    \begin{array}{c}
      X\\
      Y
    \end{array}
  \right)=\omega
  \left(
    \begin{array}{c}
      X\\
      Y
    \end{array}
  \right) \ .
  \label{RPA}
\end{eqnarray}
The nonzero matrix elements of the $2\times2$ matrices $A$ and $B$ are
\bas
  A_{11}&=& \varepsilon_1 \cos{2\alpha}+\frac{3}{2}v\sin^2{2\alpha}\ , \\
  A_{22}&=& \varepsilon_2-\varepsilon_1\sin^2{\alpha}+\frac{v}{2}\sin^2{2\alpha}\ ,\\
  B_{11}&=& -v(\cos^4\alpha+\sin^4\alpha) \ ,\\
  B_{22}&=& -v\cos^2\alpha \ .
\eas

Solving for the eigenfrequencies $\omega$ yields
\ba
  \omega_1^2&=&\left\{
    \begin{array}{l l}
      \varepsilon_1^2-v^2\ , &\varepsilon_1> v\\
      2(v^2-\varepsilon_1^2)\ , &\varepsilon_1< v
    \end{array} \right. \\
  \omega_2^2&=&\left\{ 
    \begin{array}{l l}
      \varepsilon_2^2-v^2 \ , &\varepsilon_1> v\\
      (\varepsilon_2-\varepsilon_1)(\varepsilon_2+v)\ ,& \varepsilon_1< v \ .
    \end{array} \right. 
  \label{omega}
\ea
The total energy is 
\be
E=E_{\rm HF}+ {1\over 2}\left(\sum_{i=1}^2 \omega_i-{\rm Tr} A\right) \ , 
\ee
and the result thus becomes
\ba
\label{e1}
E&=& {1\over 2} \sum_{j=1}^2 \left(\sqrt{\varepsilon_j^2-v^2}
-\varepsilon_j\right)\quad\mbox{for}\quad\varepsilon_1>v\ ,\\
\label{e2}
E&=& -{N+4\over 4}v 
+{1\over 2} \sqrt{\varepsilon_2-\varepsilon_1}\sqrt{v+\varepsilon_2}
\nonumber\\ 
&&+\sqrt{v^2-\varepsilon_1^2\over 2} 
+{2N+1\over 4}\varepsilon_1\nonumber\\
&& -{\varepsilon_2\over 2} - {N-1\over 4}{\varepsilon_1^2\over v} 
\quad\mbox{for}\quad\varepsilon_1<v \ .
\ea
In the weak-coupling regime, the energy~(\ref{e1}) is very simply
related to the energy~(\ref{erg1}) for the spherical phase, while the
energy of the strong-coupling regime differs mainly from its deformed
counterpart~(\ref{erg2}) through the term that is nonanalytical in the
level splitting $\varepsilon_2-\varepsilon_1$. It seems that the
non-analyticity is an artifact of the RPA approximation.

\begin{figure}
  \includegraphics[width=0.5\textwidth,angle=0]{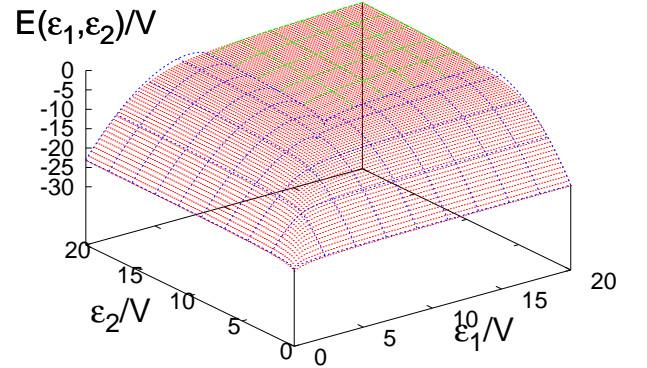}
  \caption{(color online) Ground-state energy
  $E(\varepsilon_1,\varepsilon_2)$ as a function of the
  single-particle energies $\varepsilon_{1,2}$ from the RPA (mesh of
  dashed lines) and the exact result (dots) for $N=10$ fermions in the
  three-level Lipkin model. The flat region corresponds to the
  weak-coupling regime, and the cusp of the RPA results marks the
  boundary to the strong-coupling regime.
  \label{SplitFuncFig}}
\end{figure}

Fig.~\ref{SplitFuncFig} compares this analytical
expressions~(\ref{e1}) and (\ref{e2}) with the exact energy for a
system of $N=10$ fermions. Again, the RPA fails at the boundary
between the regime of strong and weak coupling, and the energy is not a
convex function of the single-particle energies.

The energy functional results from the Legendre transform 
\be
\label{f2}
F(n_1,n_2) = E(\varepsilon_1,\varepsilon_2) - \sum_{j=1}^2 n_j
\varepsilon_j \ , 
\ee 
and it is understood that $\varepsilon_j=\varepsilon_j(n_1,n_2)$
results from the inversion of the occupation numbers 
\be 
n_j\equiv {\partial E\over \partial\varepsilon_j} \ .  
\ee
 
We start with the weak-coupling regime $\varepsilon_1 > v$ and find
\be
n_j=\frac{1}{2}\left(\frac{\varepsilon_j}
{\sqrt{\varepsilon_j^2-v^2}}-1\right) \ . 
\label{occnum1}
\ee
Thus, the occupation number of the two energy levels depends only on
the energy of the level itself, and the strength of the two-body
interaction. Note that the occupation numbers are symmetric under
exchange of level 1 and level 2.  The inversion of Eq.~(\ref{occnum1})
is straightforward, and we obtain the energy functional
\begin{equation}
\label{fsym2}
  F(n_1,n_2)=-|v |\sum_{j=1}^2\sqrt{n_j(n_j+1)} \ .
\end{equation}
Note that the functional~(\ref{fsym2}) is symmetric under exchange of
level 1 and level 2, and that it is simply the sum of two functionals.
Its form could have almost been guessed from the
functional~(\ref{fsym1}) for degenerate levels. 

\begin{figure}
  \includegraphics[width=0.5\textwidth,angle=-0]{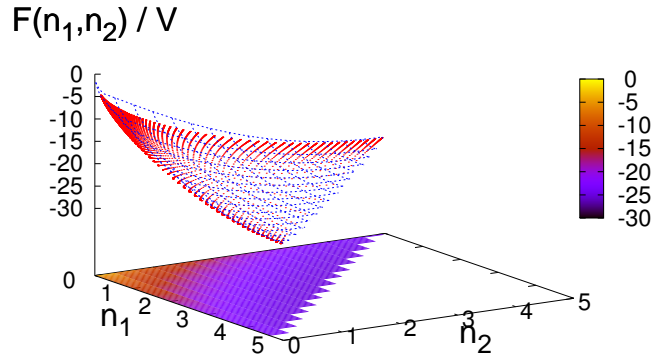}
  \caption{(color online) Energy functional $F(n_1,n_2)$ of the
  three-level Lipkin model with non-degenerate single-particle levels
  for $N=10$ fermions. The exact result is shown as dots, while the RPA
  result is shown as a mesh and the corresponding contour.
  \label{figF2}}
\end{figure}

We turn to the strong-coupling regime $\varepsilon_1<v$ which once again
proves more difficult.  The occupation numbers
\begin{eqnarray}
\label{n1}
  n_1&=&\frac{2N+1}{4}-{x\over 4}-
\frac{N-1}{2}{\varepsilon_1\over v}
-\frac{\varepsilon_1}{\sqrt{2(v^2-\varepsilon_1^2})}\\
\label{n2}
  n_2&=&\frac{1}{4}\left(x+x^{-1} -2\right) 
\end{eqnarray}
do not decouple as in the weak-coupling limit. Here, we employed the 
shorthand 
\be
\label{x}
  x\equiv\sqrt{\frac{\varepsilon_2+v}{\varepsilon_2-\varepsilon_1}}\ .
\ee
We solve Eq.~(\ref{n2}) for $x$ and obtain
\be
  x(n_2)=2 n_2+1+2\sqrt{n_2(n_2+1)} \ .
\ee
We insert this result into Eq.~(\ref{n1}), approximate
$\varepsilon_1\ll v$ and solve for $\varepsilon_1$.  This expansion 
again renders the resulting functional convex, and we obtain
\be
\label{eps1}
  \varepsilon_1(n_1,n_2)= \frac{N-2n_1+1/2-x/2}{N-1+\sqrt{2}} v \ .
\ee
Finally, we insert this result into Eq.~(\ref{x}) and find
\be
\label{eps2}
  \varepsilon_2(n_1,n_2)=\frac{v+x^2 \varepsilon_1}{x^2-1} \ .
\ee
The insertion of the expressions~(\ref{eps1}) and (\ref{eps2}) into
the functional~(\ref{f2}) thus yields the desired
expression. Figure~\ref{figF2} shows the resulting functional and
compares it to the exact solution. The agreement between the exact
result and the RPA result is quite satisfactory. Note that the
boundary between the regimes of strong and weak coupling is a
one-dimensional line in the region of small occupation numbers
$n_{1,2}<1$.

\section{Summary}

We constructed the energy functional for the three-level Lipkin model
with degenerate and non-degenerate excited levels. Our analytical
results are based on RPA calculations and subsequent Legendre
transformations. They agree well with exact numerical results in the
limits of strong and weak coupling, respectively, but the RPA fails at
the boundary between both regimes. In particular, the RPA energy is
not a convex function of the single-particle energies in the
strong-coupling regime. 

In the case of non-degenerate excited levels, a continous symmetry is
broken by single-particle terms of the Hamiltonian while the two-body
interaction alone preserves this symmetry. In the weak-coupling
regime, the energy functional obtained from non-degenerate levels is
simply related to the functional obtained for degenerate excited
levels. Both exhibit a square-root singularity for small occupation
numbers. The relationship is more complicated in the strong-coupling
regime. However, the derived results should be useful in the
construction of an occupation-number based energy functional for
nuclear masses. Recall that the occupation number-based mass formula by
Duflo and Zuker~\cite{Duf95} is superior to mass table calculations
that employ nuclear energy-density functionals~\cite{Lun03}. This
gives prospect to the development of a global nuclear energy
functional based on shell model occupations.

This research was supported in part by the U.S. Department of Energy
under Contract Nos.\ DE-FG02-96ER40963, DE-FG02-07ER41529 (University
of Tennessee) and DE-AC05-00OR22725 with UT-Battelle, LLC (Oak Ridge
National Laboratory).

\end{document}